\def\tr{{\rm tr}\,}
\def\Tr{{\rm Tr}\,}
\def\wt{\widetilde}
\def\sgn{{\rm sgn\,}}
\def\b{\bibitem}
\def\be{\begin{equation}}
\def\ee{\end{equation}}
\def\bea{\begin{eqnarray}}
\def\eea{\end{eqnarray}}
\def\bml{\begin{mathletters}}
\def\eml{\end{mathletters}}
\documentstyle[aps,prb,eqsecnum,psfig,epsf,floats]{revtex}
\draft
\begin{document}
\def\SNG{{\em Physical Review Style and Notation Guide}}
\def\LUG {{\em \LaTeX{} User's Guide \& Reference Manual}}
\def\btt#1{{\tt$\backslash$\string#1}}%
\def\REVTeX{REV\TeX}
\def\AmS{{\protect\the\textfont2
        A\kern-.1667em\lower.5ex\hbox{M}\kern-.125emS}}
\def\AmSLaTeX{\AmS-\LaTeX}
\def\BibTeX{\rm B{\sc ib}\TeX}
\twocolumn[\hsize\textwidth\columnwidth\hsize\csname@twocolumnfalse%
\endcsname

\title{Theory of Disordered Itinerant Ferromagnets II: Metal-Insulator 
       Transition
}
\author{T.R.Kirkpatrick}
\address{Institute for Physical Science and Technology, and Department of
 Physics\\
 University of Maryland,\\
 College Park, MD 20742}
\author{D.Belitz}
\address{Department of Physics and Materials Science Institute\\
University of Oregon,\\
Eugene, OR 97403}
\date{\today}
\maketitle

\begin{abstract}
The theory for disordered itinerant ferromagnets developed in a previous paper
is used to construct a simple effective field theory that is capable of 
describing the quantum phase transition from a ferromagnetic metal to a 
ferromagnetic insulator. It is shown that this transition is in the same 
universality class as the one from a paramagnetic metal to a paramagnetic 
insulator in the presence of an external magnetic field, and that strong
corrections to scaling exist in this universality class. The experimental 
consequences of these results are discussed.
\end{abstract}
]

\section{Introduction}
\label{sec:I}

\begin{figure}[ht]
\vskip 2.5cm
\epsfxsize=6.25cm
\epsffile{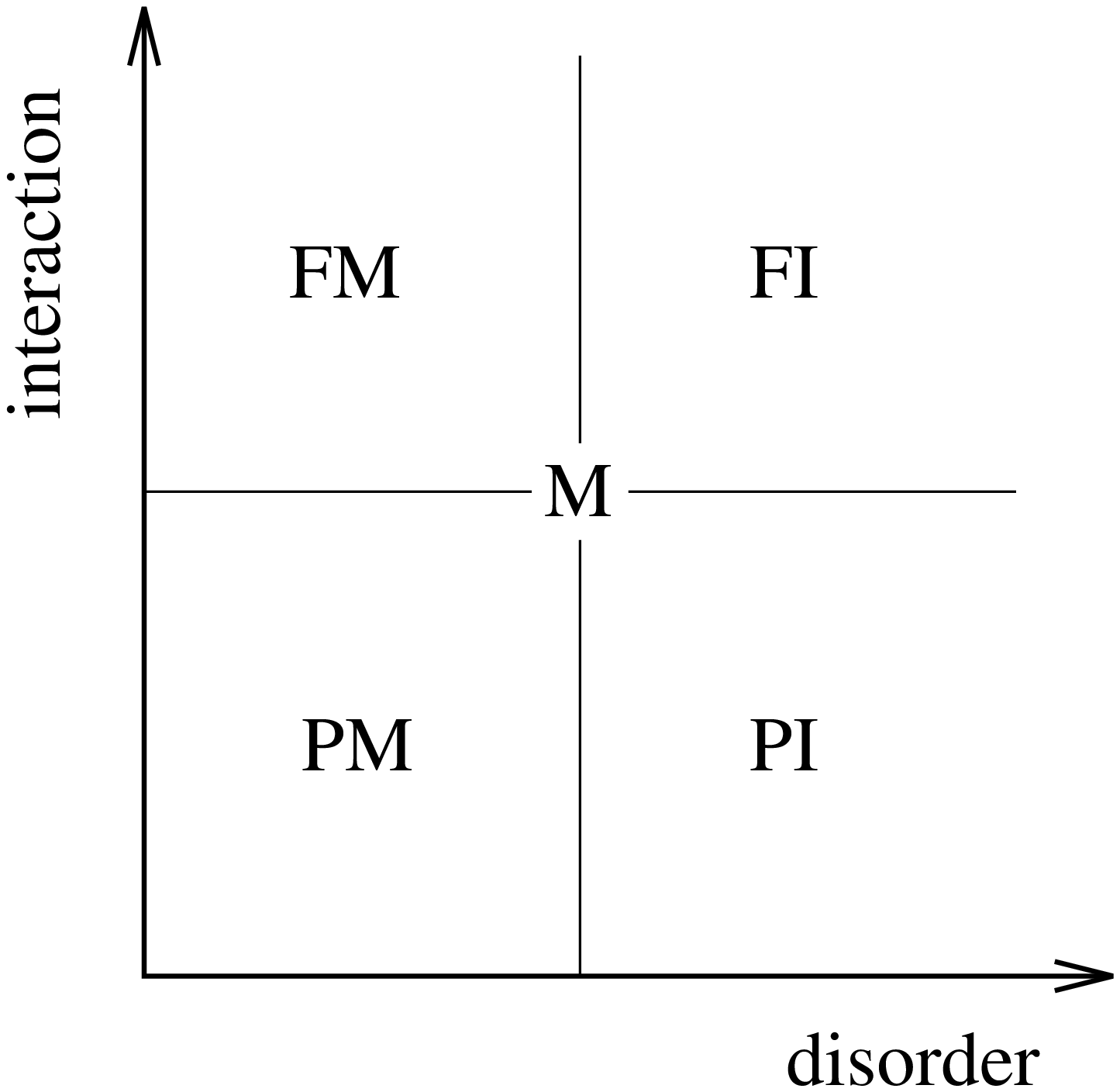}
\vskip -1.5cm
\caption{Schematic phase diagram of disordered, interacting electrons, showing
 paramagnetic metal (PM), paramagnetic insulator (PI), ferromagnetic metal (FM),
 and ferromagnetic insulator (FI) phases. M denotes a multicritical point.}
\label{fig:1}
\end{figure}

It is well known that interacting electrons in the presence of
quenched disorder at zero 
temperature form a disordered Fermi-liquid or paramagnetic metal state 
that shows, with increasing 
disorder, an instability against the formation of an insulator. This 
Anderson-Mott transition (PM to PI in the schematic phase diagram shown in
Fig.\ \ref{fig:1}) is believed to be the metal-insulator transition 
observed in doped semiconductors and other disordered electron systems, and 
it has been studied theoretically in considerable detail.\cite{F,R} Similarly,
with increasing exchange interaction, the Fermi liquid state is unstable
against the formation of long-range ferromagnetic order (PM to FM in 
Fig.\ \ref{fig:1}).
This quantum phase transition has also been studied, both with and without
quenched impurities.\cite{Hertz}
In the Fermi-liquid phase, the PM-PI transition is preceded by nonanalyticities
of various observables (e.g. the conductivity, the tunneling density of states,
the spin susceptibility, etc.) as functions of wavenumber, frequency, or 
temperature. These nonanalyticities are often referred to as ``weak-localization
effects''. They are caused by soft modes, viz. diffusive particle-hole
excitations (``diffusons''), and can be studied in perturbation 
theory.\cite{LeeRama} The diffusons are known to drive the metal-insulator
transition, at least near two-dimensions, which is the lower critical
dimensionality for this transition.
The analogous soft-mode effects in the metallic ferromagnetic 
state have recently been investigated in Ref.\ \onlinecite{us_paper_I}
(to be referred to as (I)),
but the quantum phase transition that must occur from a ferromagnetic
metal to a ferromagnetic insulator upon increasing the disorder (FM to FI in
Fig.\ \ref{fig:1}) has never been considered. 

In this paper we address the latter problem. In particular, we derive and 
analyze an effective field theory that is capable of describing the disorder 
driven transition from a ferromagnetic metal to a ferromagnetic insulator.
An interesting theoretical question that arises in this context is the
role of the Goldstone modes that occur due to the broken spin rotational
symmetry, i.e. the spin waves. Since they constitute soft modes in addition
to the diffusons, one would {\it a priori} expect them to influence the
critical behavior. It was shown in (I) that the
Goldstone modes, while they contribute to the leading frequency nonanalyticity
of $O(\Omega^{(d-2)/2})$ in the conductivity, yield a prefactor that
is of $O(1)$, while the diffusons contribute a prefactor that is of
$O(1/(d-2))$ and thus diverges as $d\rightarrow 2$.
Since it is known that this singularity drives the transition
near two-dimensions,\cite{hi-d_footnote} it follows that the Goldstone 
mode do not contribute to the asymptotic critical behavior.
Largely as a consequence of this, the 
ferromagnetic-metal-to-ferromagnetic-insulator transition turns out to be
in the same universality class as the one from a paramagnetic metal to a
paramagnetic insulator in the presence of an external magnetic 
field. We find strong corrections to scaling for this universality
class.\cite{MF_Footnote}

Another important motivation for the present study is the recently observed
apparent metal-insulator transition in Si MOSFETs and other two-dimensional
($2$-$d$) electron systems,\cite{2dMIT} which contradicts the orthodox
theoretical results that predict an insulating state in $d=2$ even for
arbitrarily weak disorder.\cite{gang_of_4,F,R}
Since it is known that magnetic fluctuations have a tendency to increase the
conductivity in or close to two-dimensions,\cite{R,Runaway_Footnote} 
it is conceivable that
there might be a ferromagnetic metallic phase at small but nonzero disorder
in $d=2$. We find that this is not the case, which rules out a possible
mechanism for a metal-insulator transition in $d=2$.

This paper is organized as follows. In Sec.\ \ref{sec:II} we briefly recall the
general $Q$-matrix field theory for itinerant ferromagnets that
was developed in (I). On the basis of this, and using the perturbative results 
of (I), we construct an effective theory for the most relevant soft modes in 
the system. This theory takes the form of a generalized nonlinear $\sigma$ 
model. In Sec.\ \ref{sec:III}
we show that this model describes a 
ferromagnetic-metal-to-ferromagnetic-insulator transition in $d>2$, and
we calculate the critical behavior at this transition in a $d=2+\epsilon$
expansion. In Sec.\ \ref{sec:IV} we conclude with a discussion of our
results.

\section{Effective field theory for disordered itinerant ferromagnets}
\label{sec:II}

\subsection{$Q$-matrix theory}
\label{subsec:II.A}

In (I) it was shown that disordered itinerant ferromagnets are described
by the following action,
\begin{mathletters}
\label{eqs:2.1}
\begin{eqnarray}
{\cal A}[Q,{\wt\Lambda}] &=& {\cal A}_{\rm dis} + {\cal A}_{\rm int}
                           + \frac{1}{2}\,\Tr\ln\left(G_0^{-1} - i{\wt\Lambda}
                                       \right)
\nonumber\\
  && + \int d{\bf x}\ \tr\left({\wt\Lambda}({\bf x})\,Q({\bf x})\right)\quad.
\label{eq:2.1a}
\end{eqnarray}
Here
\begin{equation}
G_0^{-1} = -\partial_{\tau} + \partial_{\bf x}^2/2m_{\rm e} + \mu\quad,
\label{eq:2.1b}
\end{equation}
is the inverse free electron Green operator, with $\partial_{\tau}$ and
$\partial_{\bf x}$ derivatives with respect to imaginary time and position,
respectively, $m_{\rm e}$ is the electron mass, 
and $\mu$ the chemical potential.
$Q$ and $\wt\Lambda$ are matrix fields that carry two Matsubara frequency
indices $n$ and $m$, and two replica indices $\alpha$ and $\beta$. The
matrix elements $Q_{nm}^{\alpha\beta}$ and ${\wt\Lambda}_{nm}^{\alpha\beta}$
are spin-quaternion valued. They are conveniently expanded in a basis
\be
Q_{nm}^{\alpha\beta}({\bf x}) = \sum_{r,i=0}^{3} (\tau_r\otimes s_i)\,
   {^i_rQ_{nm}^{\alpha\beta}}({\bf x}) \quad,
\label{eq:2.1c}
\ee
\eml%
and analogously for $\wt\Lambda$. Here
$\tau_0 = s_0 = \openone_2$ is the
$2\times 2$ unit matrix, and $\tau_j = -s_j = -i\sigma_j$, $(j=1,2,3)$,
with $\sigma_{1,2,3}$ the Pauli matrices. In this basis, $i=0$ and $i=1,2,3$
describe the spin singlet and the spin triplet, respectively. An explicit
calculation reveals that $r=0,3$ corresponds to the particle-hole channel
while $r=1,2$ describes the particle-particle channel.
In Eq.\ (\ref{eq:2.1a}),
$\Tr$ denotes a trace over all degrees of freedom, including the continuous
position variable, while $\tr$ is a trace over all those discrete indices that
are not explicitly shown. 
For the disorder part of the action one finds\cite{us_fermions}
\be
{\cal A}_{\rm dis}[Q] = \frac{1}{\pi N_F\tau}\int d{\bf x}\
                       \tr \bigl(Q({\bf x})\bigr)^2\quad.
\label{eq:2.2}
\ee
with $\tau$ the single-particle scattering or relaxation 
time.\cite{dis_footnote}
The electron-electron interaction 
${\cal A}_{\rm int}$ is conveniently decomposed into four pieces 
that describe the interaction
in the particle-hole and particle-particle spin-singlet and spin-triplet 
channels.\cite{us_fermions} We will need only the particle-hole channel,
and thus write
\begin{mathletters}
\label{eqs:2.3}
\begin{equation}
{\cal A}_{\rm int}[Q] = {\cal A}_{\rm int}^{\,(s)} 
     + {\cal A}_{\rm int}^{\,(t)}\quad,
\label{eq:2.3a}
\end{equation}
\begin{eqnarray}
{\cal A}_{\rm int}^{\,(s)}&=&\frac{T\Gamma^{(s)}}{2}\int d{\bf x}
              \sum_{r=0,3}(-1)^r \sum_{n_1,n_2,m}\sum_\alpha
\nonumber\\
&&\times\left[\tr \left((\tau_r\otimes s_0)\,Q_{n_1,n_1+m}^{\alpha\alpha}
({\bf x})\right)\right]
\nonumber\\
&&\times\left[\tr \left((\tau_r\otimes s_0)\,Q_{n_2+m,n_2}^{\alpha\alpha}
({\bf x})\right)\right]\quad,
\label{eq:2.3b}
\end{eqnarray}
\begin{eqnarray}
{\cal A}_{\rm int}^{\,(t)}&=&\frac{T\Gamma^{(t)}}{2}\int d{\bf x}
    \sum_{r=0,3}(-1)^r \sum_{n_1,n_2,m}\sum_\alpha\sum_{i=1}^3
\nonumber\\
&&\times\left[\tr\left((\tau_r\otimes s_i)\,Q_{n_1,n_1+m}^{\alpha\alpha}
({\bf x})\right)\right]
\nonumber\\
&&\times\left[\tr\left((\tau_r\otimes s_i)\,Q_{n_2+m,n_2}^{\alpha\alpha}
({\bf x})\right)\right]\quad.
\label{eq:2.3c}
\end{eqnarray}
\eml%
Here $\Gamma^{(s)}>0$ and $\Gamma^{(t)}>0$ are the spin-singlet and 
spin-triplet interaction amplitudes, respectively. $\Gamma^{(t)}$ is
responsible for producing magnetism.

As was shown in (I), the Goldstone modes do not contribute
leading singular terms to the conductivity, or to any of the two-point vertices,
in perturbation theory up to one-loop order as $d\rightarrow 2$ in the metallic
ferromagnetic phase. It was further shown that they
do not contribute to the renormalization of the DOS, and hence they do
not contribute to the wave function renormalization. It follows that both
the heat and the charge diffusion constants do not carry any singular 
renormalizations due to Goldstone modes. If we assume that this
signals the absence of Goldstone mode effects on the critical properties
near the metal-insulator transition as well,
we can ignore the Goldstone
modes for the purpose of constructing an effective field theory for the
soft modes that drive the metal-insulator transition.\cite{hi-d_footnote_2}
Furthermore, we can ignore the particle-particle channel, which is massive
in a system with a nonvanishing magnetization. Accordingly, we drop
both the particle-particle channel ($r=1,2$) and the transverse spin-triplet
channels ($i=1,2$) from our model definition. For the remaining soft modes,we 
will now construct an effective theory
by generalizing the procedure followed in Ref.\ \onlinecite{us_fermions}.

\subsection{Soft and massive modes}
\label{subsec:II.B}

Let us briefly recall the basic philosophy behind the derivation of a
nonlinear $\sigma$-model in Ref.\ \onlinecite{us_fermions}, which in
turn was based on the work by Sch{\"a}fer and Wegner\cite{SchaferWegner}
on noninteracting electrons. First one realizes, by means of a Ward
identity, that the soft modes are given by the matrix elements
$Q_{nm}$ with $nm<0$, while the $Q_{nm}$ with $nm>0$ are massive.
This remains true in the present case except for the Goldstone modes,
which we can neglect for our purposes. Next one block-diagonalizes
the matrix $Q$ in frequency space. Algebraic arguments show
that the most general $Q$ can be written as
\bml
\label{eqs:2.4}
\be
Q = {\cal S}P{\cal S}^{-1}\quad,
\label{eq:2.4a}
\ee
Here $\cal S$ is a matrix that represents an element of the coset space
${\rm USp}(8Nn,{\cal C})/{\rm USp}(4Nn,{\cal C})\times {\rm USp}(4Nn,{\cal C})$,
and $P$ is block-diagonal in Matsubara frequency space
\be
P = \left(\matrix{P^> & 0   \cr
                  0   & P^< \cr}\right)\quad,
\label{eq:2.4b}
\ee
\eml%
where $P^>$ and $P^<$ are matrices with elements $P_{nm}$ where
$n,m>0$ and $n,m<0$, respectively. It is further convenient to define
a transformed field $\Lambda$ by
\be
\Lambda({\bf x}) = {\cal S}^{-1}({\bf x}){\wt\Lambda}({\bf x}){\cal S}({\bf x})
   \quad,
\label{eq:2.5}
\ee
and to write the action in terms of these variables,
\bea
{\cal A}\left[{\cal S},P,\Lambda\right]&=&{\cal A}_{\rm dis}[P]
            + {\cal A}_{\rm int}[{\cal S}P{\cal S}^{-1}]
\nonumber\\
&&+ \frac{1}{2}\,\Tr\ln \left( G_0^{-1} - i{\cal S}\Lambda {\cal S}^{-1}\right)
\nonumber\\
&&+ \int d{\bf x}\,\tr\bigl(\Lambda({\bf x}) P({\bf x})\bigr)\ ,
\label{eq:2.6}
\eea

The next step is to expand $\cal S$, $P$, and $\Lambda$ about their
respective saddle-point values, which we denote by $\langle{\cal S}\rangle$,
$\langle P\rangle$, and $\langle\Lambda\rangle$, respectively. From
Sec.\ \ref{subsec:II.B} in (I) we have
\bml
\label{eqs:2.7}
\be
\langle{\cal S}\rangle = \openone\otimes\tau_0\quad,
\label{eq:2.7a}
\ee
\bea
\langle P\rangle_{12} &=& \delta_{12}\ \frac{i}{2V}\sum_{\bf p} \left[\right.
   (\tau_0\otimes s_0)\ {\cal G}_{n_1}({\bf p}) 
\nonumber\\
&&\qquad + (\tau_3\otimes s_3)\ {\cal F}_{n_1}({\bf p})\left.\right]\ ,
\label{eq:2.7b}
\eea
\bea
\langle\Lambda\rangle_{12}&=&\delta_{12}\ (\tau_0\otimes s_0)\ 
     \frac{-i}{\pi N_{\rm F}\tau}\ \frac{1}{V}
                  \sum_{\bf p}{\cal G}_{n_1}({\bf p})
\nonumber\\
&&\hskip -1pt -\ \delta_{12}\ (\tau_0\otimes s_0)\ 2i\Gamma^{(s)}T\sum_{m}
   e^{i\omega_m 0}\ \frac{1}{V}\sum_{\bf p}{\cal G}_m({\bf p})
\nonumber\\
&&\hskip -1pt +\ \delta_{12}\ (\tau_3\otimes s_3)\ 
   \frac{-i}{\pi N_{\rm F}\tau}\ 
     \frac{1}{V}\sum_{\bf p}{\cal F}_{n_1}({\bf p})
\nonumber\\
&&\hskip -1pt +\ \delta_{12}\ (\tau_3\otimes s_3)\ 2i\Gamma^{(t)}T\sum_{m}
   e^{i\omega_m 0}\ \frac{1}{V}\sum_{\bf p}{\cal F}_m({\bf p})\ ,
\nonumber\\
\label{eq:2.7c}
\eea
\eml%
with ${\cal G}$ and ${\cal F}$ from (I) Eqs.\ (2.13).
In the popular approximation that replaces the wavevector sum over a Green
function by an integral over $\xi_{\bf p} = {\bf p}^2/2m_{\rm e}$,\cite{AGD} 
we have
\bml
\label{eqs:2.8}
\bea
\frac{1}{V}\sum_{\bf p}{\cal G}_n({\bf p})&\approx&\frac{-i\pi}{2}\,N_{\rm F}
   \,\sgn \omega_n\quad,
\label{eq:2.8a}\\
\frac{1}{V}\sum_{\bf p}{\cal F}_n({\bf p})&\approx&0\quad.
\label{eq:2.8b}
\eea
\eml%
In this approximation,\cite{approx_footnote} 
we can write Eq.\ (\ref{eq:2.7b}) as
\bml
\label{eqs:2.9}
\be
\langle P\rangle_{12} \approx \pi_{12} \quad,
\ee
with
\be
\pi_{12} = \delta_{12}\,(\tau_0\otimes s_0)\,\sgn\omega_{n_1}\quad,
\label{eq:2.9b}
\ee
\eml%
with $\omega_n = 2\pi T(n+1/2)$ a fermionic Matsubara frequency.
For our purposes this approximation will be sufficient for reasons
that were explained in detail in Ref.\ \onlinecite{us_fermions}.

We now write 
\be
P = \langle P\rangle + \Delta P\quad,\quad\Lambda = \langle\Lambda\rangle
    + \Delta\Lambda\quad,
\label{2.10}
\ee
and expand in powers of $\Delta P$, $\Delta\Lambda$, and derivatives of
${\cal S}$. Let us first consider the $\Tr\ln$ term in Eq.\ (\ref{eq:2.6}).
Using the cyclic property of the trace, we can write it in the form
\bml
\label{eqs:2.11}
\bea
\Tr\ln (G_0^{-1} - {\cal S}\,i\Lambda\,{\cal S}^{-1}) &=&
   \Tr\ln ({\cal S}^{-1}\,G_0^{-1} {\cal S} - i\Lambda)
\nonumber\\
&&\hskip -90pt = \Tr\ln (G_{\rm sp})^{-1}) + \Tr\ln\left[\right. 1 +
G_{\rm sp}\,{\cal S}^{-1}(\partial_{\tau}{\cal S})
\nonumber\\
&&\hskip -75pt + \frac{1}{m}\,G_{\rm sp}\,{\cal S}^{-1}\,(\nabla{\cal S})\,\nabla
     +\frac{1}{2m}\,G_{\rm sp}\,{\cal S}^{-1}(\nabla^2{\cal S})
\nonumber\\
   && - G_{\rm sp}\,i(\Delta\Lambda)\left.\right]\quad.
\label{eq:2.11a}
\eea
with
\be
G_{\rm sp} = \left(G_0^{-1} -i\langle\Lambda\rangle\right)^{-1}\quad,
\label{eq:2.11b}
\ee
\eml%
the saddle-point Green function.
This is formally the same expression as in the absence of
ferromagnetism,\cite{us_fermions} only the saddle-point Green function is
more complicated. In particular, the transformation matrix ${\cal S}$ appears
only in conjunction with some derivative and is therefore soft, while the 
fluctuations $\Delta\Lambda$ are massive. Expanding the second term on the
right-hand side, the simplest contribution is the one involving the time
derivative,
\bea
\Tr\left[G_{\rm sp}\,{\cal S}^{-1}\,(\partial_{\tau}\,{\cal S})\right] &&
\nonumber\\
&&\hskip -40pt =
   \int d{\bf x}\ \tr\left[i\Omega\,{\cal S}({\bf x})\,G_{\rm sp}({\bf x}=0)\,
     {\cal S}^{-1}({\bf x})\right]
\nonumber\\
&&\hskip -40pt = \frac{\pi N_{\rm F}}{2} \int d{\bf x}\ 
   \tr [\Omega\,{\hat Q}({\bf x})] + O(\Omega^2\,{\hat Q})\ .
\label{eq:2.12}
\eea
Here
\bml
\label{eqs:2.13}
\be
\Omega_{12} = (\tau_0\otimes s_0)\,\delta_{12}\,\Omega_{n_1}\quad,
\label{eq:2.13a}
\ee
is a frequency matrix with $\Omega_n = 2\pi Tn$ a bosonic Matsubara
frequency, and
\be
{\hat Q}({\bf x}) = {\cal S}({\bf x})\,\pi\,{\cal S}^{-1}({\bf x})\quad,
\label{eq:2.13b}
\ee
\eml%
with $\pi$ from Eq.\ (\ref{eq:2.9b}). Here we have made use of
Eqs.\ (\ref{eqs:2.8}).\cite{integrals_footnote} This is the same result as the
one obtained in the absence of ferromagnetism.\cite{us_fermions}

We now turn to the gradient terms. It is convenient to define a matrix valued
$d$-dimensional vector field
\be
{\bf s}({\bf x}) =  {\cal S}^{-1}({\bf x})\,(\nabla{\cal S})({\bf x})\quad,
\label{eq:2.14}
\ee
and to expand in powers of ${\bf s}$. The term linear in ${\bf s}$ vanishes
for symmetry reasons. To $O({\bf s}^2)$, both the next-to-last term on the
right-hand side of Eq.\ (\ref{eq:2.11a}) and the square of the preceding term
contribute. So far our gradient expansion has been completely general. In
order to evaluate the terms of $O({\bf s}^2)$, we now remember that we can
neglect the spin waves for the purpose of deriving a soft-mode transport
theory, i.e. we have dropped the channels $r=0,3$, $i=1,2$ in the spin quaternion
expansion, Eq.\ (\ref{eq:2.1c}). Furthermore, it is well known that the Cooper
channel ($r=1,2$) is massive in an external magnetic field,\cite{LeeRama,R}
and the same is true in a ferromagnetic state. Consequently, the only
spin-quaternion degrees of freedom present in ${\bf s}_{12}$ are $\tau_{0,3}$
and $s_{0,3}$, and ${\bf s}_{12}$ commutes with $\tau_3\otimes s_3$.
This simplifies the evaluation of the gradient terms substantially, and we
obtain
\bml
\label{eqs:2.15}
\bea
\Tr G_{\rm sp}\,{\cal S}^{-1}(\nabla^2\,{\cal S}) - \frac{1}{m}\,\Tr
  (G_{\rm sp}\,{\cal S}^{-1}\,(\nabla{\cal S})\nabla)^2 =
\nonumber\\
\sum_{12}\sum_{\bf q}\int d{\bf x}\Bigl[\eta_{12,ij}^s({\bf q})\,
   \tr\left(s_{12}^i({\bf q})\,s_{21}^j({\bf -q})\right)
\nonumber\\
+\eta_{12,ij}^a({\bf q})\,\tr\left((\tau_3\otimes s_3)\,s_{12}^i({\bf q})\,
      s_{21}^j({\bf -q})\right)\Bigr]\quad,
\label{eq:2.15a}
\eea
where
\be
\eta^s = (\eta^+ + \eta^-)/2\quad,\quad\eta^a = (\eta^+ - \eta^-)/2\quad,
\label{eq:2.15b}
\ee
with
\bea
\eta_{12,ij}^{\pm}({\bf q}) &=& \delta_{ij}\,\frac{1}{2}\,
   \left[{\cal G}^{\pm}_{n_1}({\bf q}) + {\cal G}^{\pm}_{n_2}({\bf q})\right]
\nonumber\\
&&+ \frac{1}{m}\,q_i q_j\,{\cal G}^{\pm}_{n_1}({\bf q})\,
     {\cal G}^{\pm}_{n_2}({\bf q})\quad,
\label{eq:2.15c}
\eea
\eml%
with ${\cal G}_n^{\pm}$ the Green functions defined in (I) Eq.\ (2.13c). 
Equations\ (\ref{eqs:2.15}) are
generalizations of the corresponding expressions in the absence of
ferromagnetism.\cite{us_fermions}

\subsection{The nonlinear $\sigma$ model}
\label{subsec:II.C}

The remaining steps in the derivation of an effective field theory proceed
in analogy to Ref.\ \onlinecite{us_fermions}. In particular, we integrate
out the massive modes $P$ and $\Lambda$ in tree approximation, i.e. we
neglect all fluctuations $\Delta P$ and $\Delta\Lambda$. $\eta^{\pm}$
can be related to the conductivity in self-consistent Born approximation of
a system whose chemical potential has been shifted 
from its value for nonmagnetic electrons by 
$\pm\Delta=\pm\Gamma^{(t)} M/\mu_{\rm B}$. Here $M$ is the magnetization in
Stoner approximation, and $\mu_{\rm B}$ is the Bohr magneton (see (I) Eq.\ 
(2.15)). Denoting these conductivities by $\sigma_0^{\pm}$, and defining 
the bare coupling constants
\bml
\label{eqs:2.16}
\bea
1/G &=& \frac{\pi}{4}\,m\,(\sigma_0^+ + \sigma_0^-)\quad,
\label{eq:2.16a}\\
1/G_3 &=& \frac{\pi}{4}\,m\,(\sigma_0^+ - \sigma_0^-)\quad,
\label{eq:2.16b}\\
H &=& \pi\,N_{\rm F}/8\quad,
\label{eq:2.16c}
\eea
\eml%
we obtain for the effective action
\bea
{\wt{\cal A}}&=&\frac{-1}{2G}\int d{\bf x}\ \tr[\nabla{\wt Q}({\bf x})]^2
                + 2H\int d{\bf x}\ \tr [\Omega\,{\wt Q}({\bf x})] 
\nonumber\\
&&- \frac{1}{2G_3}\int d{\bf x}\ \tr\left((\tau_3\otimes s_3)
         [\nabla{\wt Q}({\bf x})]^2\right)+ {\cal A}_{\rm int}[{\wt Q}]
     \quad.
\nonumber\\
\label{eq:2.17}
\eea
Here ${\wt Q} = {\hat Q} - \pi$ with $\pi$ the matrix defined in 
Eq.\ (\ref{eq:2.9b})
and ${\cal A}_{\rm int}$ from Eqs.\ (\ref{eqs:2.3}). We recognize this action
as the generalized nonlinear $\sigma$ model for disordered interacting 
electrons,\cite{F} augmented by the term with
coupling constant $1/G_3$ that is proportional to the magnetization.

It turns out that the bare action, Eq.\ (\ref{eq:2.17}), is not sufficient to
completely describe the effects of magnetic long-range order, even if one
ignores the spin waves as we did. As we will see, under renormalization two 
additional terms are generated. One is a frequency coupling that is analogous
to the second gradient term in Eq.\ (\ref{eq:2.17}), and the other is an
electron-electron interaction term that is not present in nonmagnetic systems.
We therefore need to add these terms to our action. Denoting the respective 
coupling constants by $H_3$ and $K_3$, we obtain our final result for the
effective action,
\bml
\label{eqs:2.18}
\bea
{\cal A} &=& \frac{-1}{2G}\int d{\bf x}\ \tr[\nabla{\wt Q}({\bf x})]^2
                + 2H\int d{\bf x}\ \tr [\Omega\,{\wt Q}({\bf x})] 
\nonumber\\
&&\hskip -7pt - \frac{1}{2G_3}\int d{\bf x}\ \tr\left((\tau_3\otimes s_3)
         [\nabla{\wt Q}({\bf x})]^2\right)
\nonumber\\
&&\hskip -7pt + 2H_3\int d{\bf x}\ \tr\left[(\tau_3\otimes s_3)\,\Omega\,
   {\wt Q}({\bf x}) \right] + {\cal A}_{\rm int}[{\wt Q}]\ . 
\label{eq:2.18a}
\eea
Here ${\cal A}_{\rm int}$ is given by Eqs.\ (\ref{eqs:2.3}) plus the extra
term. Introducing new interaction amplitudes $K_s = -2\pi\Gamma^{(s)}$
and $K_t = 2\pi\Gamma^{(t)}$ to comform with notation used earlier,\cite{R}
we write
\be
{\cal A}_{\rm int}[Q] = {\cal A}_{\rm int}^{(s)}[Q] 
                         + {\cal A}_{\rm int}^{(t)}[Q] 
                         + {\cal A}_{\rm int}^{(3)}[Q]\quad,
\label{eq:2.18b}
\ee
\bea
{\cal A}_{\rm int}^{(s)}[Q]&=& \frac{-\pi T}{4}\,K_s\int d{\bf x}\ \sum_{1234}
   \delta_{\alpha_1\alpha_2}\,\delta_{\alpha_1\alpha_3}\,\delta_{1-2,4-3}\,
\nonumber\\
\hskip -20pt &&\times\sum_r (-)^r \tr\left[(\tau_r\otimes s_0)\,Q_{12}({\bf x})
   \right]\ 
\nonumber\\
\hskip -20pt &&\times\tr\left[(\tau_r\otimes s_0)\,Q_{34}({\bf x})\right]\quad,
\label{eq:2.18c}
\eea
\bea
{\cal A}_{\rm int}^{(t)}[Q]&=& \frac{\pi T}{4}\,K_t\int d{\bf x}\ \sum_{1234}
   \delta_{\alpha_1\alpha_2}\,\delta_{\alpha_1\alpha_3}\,\delta_{1-2,4-3}\,
\nonumber\\
&&\hskip -20pt\times\sum_r (-)^r \tr\left[(\tau_r\otimes s_3)\,Q_{12}({\bf x})
    \right]\ 
\nonumber\\
 &&\hskip -20pt\times\tr\left[(\tau_r\otimes s_3)\,Q_{34}({\bf x})\right]\quad,
\label{eq:2.18d}
\eea
\bea
{\cal A}_{\rm int}^{(3)}[Q]&=& -4\pi TK_3\int d{\bf x}\ \sum_{1234}
   \delta_{\alpha_1\alpha_2}\,\delta_{\alpha_1\alpha_3}\,\delta_{1-2,4-3}\,
\nonumber\\
&& \times\sum_{rs}\sum_{ij} m_{rs,ij}\ {^i_r Q}_{12}({\bf x})\,{^j_s Q}_{34}
   ({\bf x}) \quad,
\label{eq:2.18e}
\eea
where
\be
m_{rs,ij} = \frac{1}{4}\,\tr (\tau_3\tau_r\tau_s^{\dagger})\ 
                         \tr (s_3 s_i s_j^{\dagger})\quad.
\label{eq:2.18f}
\ee
\eml%
(This is the matrix that was denoted by $m^{03}$ in (I).)
Finally, we note that ${\hat Q}$ as defined in Eq.\ (\ref{eq:2.13b}) obeys
\be
{\hat Q}^2({\bf x}) \equiv 1\quad,\quad
{\hat Q}^{\dagger} = {\hat Q}\quad,\quad
\tr{\hat Q}({\bf x}) \equiv 0\quad.
\label{eq:2.19}
\ee
Equations (\ref{eqs:2.18},\ref{eq:2.19}) represent the analog for itinerant
ferromagnets of the nonlinear $\sigma$ model\cite{F} for paramagnetic electron
systems.

\subsection{The metallic fixed point}
\label{subsec:II.D}

Before we proceed to use the $\sigma$ model to study the quantum phase
transition from a ferromagnetic metal to a ferromagnetic insulator, let
us ascertain that the model, with some correction terms, actually describes
a metallic ferromagnetic phase in some parts of parameter space. Since
we will want to approach the transition from this phase, its existence
within the model is obviously a necessary condition for our program to
be viable.

This task is very simple, since it proceeds in exact analogy
to the demonstration in Ref.\ \onlinecite{us_fermions} that the model
with $1/G_3 = H_3 = K_3 = 0$ has a stable Fermi-liquid fixed point.
This is because the power counting procedure used to prove the existence of a
stable fixed point does not depend on structural details like the presence
of extra $\tau$ and $s$ matrices in the various terms of the action, while
such details are the only difference between the current model and the
one considered in Ref.\ \onlinecite{us_fermions}. Accordingly, we
parameterize ${\hat Q}$ in terms of a matrix $q$ with elements $q_{nm}$
whose frequency labels are restricted to $n\geq 0$, $m<0$,
\be
{\hat Q} = \left(\matrix{\sqrt{1-qq^{\dagger}} & q \cr
                 q^{\dagger} & -\sqrt{1-q^{\dagger}q}\cr}\right)\quad,
\label{eq:2.20}
\ee
and expand ${\cal S}$ in powers of $q$,
\be
{\cal S} = \openone\otimes\tau_0 + \frac{1}{2}\left(\matrix{0 & -q \cr
                                                   q^{\dagger} & 0 \cr}\right)
  + O(q^2)\quad.
\label{eq:2.21}
\ee
As in Ref.\ \onlinecite{us_fermions}, we assign scale dimensions
to $q({\bf x})$,
\bml
\label{eqs:2.22}
\be
[q({\bf x})] = (d-2)/2\quad,
\label{eq:2.22a}
\ee
and to the fluctuations of the fields $P$ and $\Lambda$,
\be
[\Delta P({\bf x})] = [\Delta\Lambda ({\bf x})] = d/2\quad.
\label{eq:2.22b}
\ee
\eml%
Here the scale dimensions $[\ldots]$ are defined such that the scale
dimension of a length $L$ is $[L]=-1$. The fixed point action then
consists of the nonlinear $\sigma$ model action, Eq.\ (\ref{eq:2.18a}),
expanded to $O(q^2)$, plus the corrections bilinear in $\Delta P$ and
$\Delta\Lambda$ that arise from Eq.\ (\ref{eq:2.6}). All other terms
are irrelevant by power counting. The arguments showing this are exactly 
the same as the ones given in Ref.\ \onlinecite{us_fermions} and need not be
repeated here. The correlation functions for this
Gaussian action are simply related to the Gaussian propagators of
Sec.\ III in (I). We will explicitly determine them in Sec.\ \ref{sec:III}
below. This will show that the fixed point action really describes
a disordered itinerant ferromagnet.

In contrast to Ref.\ \onlinecite{us_fermions}, however, we cannot
discuss the leading corrections to scaling near the stable metallic
fixed point within our current framework. The reason is our having 
neglected the transverse
spin-triplet channel that contains the Goldstone modes. While
the latter are not expected to influence the leading scaling behavior
at the critical fixed point for the reasons pointed out above, they
do contribute to the corrections to scaling near the metallic fixed
point, as indicated by their contribution to the leading nonanalytic
frequency dependence of the conductivity that was studied in (I).

\section{Metal-insulator transition on the background of ferromagnetism}
\label{sec:III}

In this section we perform a one-loop renormalization of the nonlinear
$\sigma$ model, Eqs.\ (\ref{eqs:2.18}). We first do this for general
parameter values, which leads to rather complicated flow equations.
They contain a fixed point that corresponds to the known critical fixed point
for nonmagnetic electrons in an external magnetic field.\cite{MF_Footnote} 
We then linearize about this fixed point and show that it is perturbatively
stable with respect to the additional terms in the action that represent
the presence of a nonzero magnetization.

\subsection{Parametrization, and Gaussian order}
\label{subsec:III.A}

In order to set up a loop expansion we use the parameterization for
the matrix ${\hat Q}$ that is given by Eq.\ (\ref{eq:2.20}).
Note that this parameterization builds in the constraints given in
Eq.\ (\ref{eq:2.19}). It is the matrix analog of the usual $(\sigma,{\vec\pi})$
parameterization of the $O(N)$ vector nonlinear $\sigma$ model.\cite{ZJ}
The loop expansion now proceeds as an expansion in powers of $q$. To
Gaussian order, we obtain
\bml
\label{eqs:3.1}
\be
{\cal A}^{(0)} = \frac{-4}{V}\sum_{\bf p}\sum_{1234}{^i_rq}_{12}({\bf p})\,
   {^{ij}_{rs}M}_{12,34}({\bf p})\,{^j_sq}_{34}(-{\bf p})\quad,
\label{eq:3.1a}
\ee
where the Gaussian vertex is given by
\bea
{^{ij}_{rs}M}_{12,34}({\bf p})&=&\delta_{13}\,\delta_{24}\,
   {^{ij}_{rs}M}_{12}^{(0)}({\bf p}) 
\nonumber\\
&&+ \delta_{1-2,3-4}\,\delta_{\alpha_1\alpha_2}\,\delta_{\alpha_1\alpha_3}
           \,2\pi T\,K_{rs,ij}\ ,
\label{eq:3.1b}
\eea
with
\bea
{^{ij}_{rs}M}_{12}^{(0)}({\bf p})&=&\delta_{rs}\,\delta_{ij}\,\frac{1}{G}\,
   \left({\bf p}^2 + GH\Omega_{n_1-n_2}\right) 
\nonumber\\
&&+ m_{rs,ij}\,\frac{1}{G_3}\,
      \left({\bf p}^2 + G_3H_3\Omega_{n_1-n_2}\right)\quad,
\label{eq:3.1c}
\eea
and
\be
K_{rs,ij} = \delta_{rs}\,\delta_{ij}\,\left(\delta_{i0}K_s + \delta_{i3}K_t
   \right) + m_{rs,ij}\,K_3\quad,
\label{eq:3.1d}
\ee
\eml%
with $m_{rs,ij}$ from Eq.\ (\ref{eq:2.18f}).

The Gaussian propagator can be determined by the same methods that were
employed in Sec. III of (I). We find
\bml
\label{eqs:3.2}
\be
\langle{^i_rq}_{12}({\bf k})\,{^j_sq}_{34}({\bf p})\rangle^{(0)} =
   \frac{1}{8}\,\delta({\bf k}+{\bf p})\,{^{ij}_{rs}M}^{-1}_{12,34}({\bf p})
   \quad,
\label{eq:3.2a}
\ee
where $M^{-1}$ has the structure
\bea
{^{ij}_{rs}M}^{-1}_{12,34}({\bf p})&=&\delta_{13}\,\delta_{34}\,\bigl[
   \delta_{rs}\,\delta_{ij}\,A_{n_1-n_2}({\bf p})
\nonumber\\
&&\qquad\qquad + m_{rs,ij}\,B_{n_1-n_2}({\bf p})\bigr]
\nonumber\\
&&+\delta_{1-2,3-4}\,\delta_{\alpha_1\alpha_2}\,\delta_{\alpha_1\alpha_3}\,
   \bigl[\delta_{rs}\,\delta_{ij}\,C^{i}_{n_1-n_2}({\bf p})
\nonumber\\
&&\qquad\qquad + m_{rs,ij}\,D_{n_1-n_2}({\bf p})\bigr]\quad.
\label{eq:3.2b}
\eea
\eml%
To specify the propagators $A$, $B$, $C^0\equiv C^s$, $C^{1,2,3}\equiv C^t$, 
and $D$, we define 
\bml
\label{eqs:3.3}
\bea
a \equiv a_n({\bf p}) = ({\bf p}^2 + GH\Omega_n)/G\quad,
\label{eq:3.3a}\\
b \equiv b_n({\bf p}) = ({\bf p}^2 + G_3H_3\Omega_n)/G_3\quad,
\label{eq:3.3b}
\eea
and
\bea
N&\equiv&N_n({\bf p}) = (a^2-b^2)\bigl[a^2-b^2-2bK_3\Omega_n
\nonumber\\
&&+ a(K_s+K_t)\Omega_n
   - K_3^2\Omega_n^2 + K_sK_t\Omega_n^2\bigr]\quad.
\label{eq:3.3c}
\eea
\eml%
In terms of these quantities, we have
\bml
\label{eqs:3.4}
\bea
A_n({\bf p})&=&a/(a^2-b^2)\quad,
\label{eq:3.4a}\\
B_n({\bf p})&=&-b/(a^2+b^2)\quad,
\label{eq:3.4b}\\
C_n^s({\bf p})&=&\frac{-2\pi T}{N}\,\bigl[a^2K_s + b^2K_t 
               + a(-2bK_3 - K_3^2\Omega_n
\nonumber\\
&&\qquad\qquad\qquad   + K_sK_t\Omega_n)\bigr]\quad,
\label{eq:3.4c}\\
C_n^t({\bf p})&=&\frac{-2\pi T}{N}\,\bigl[a^2K_t + b^2K_s 
                   + a(-2bK_3 - K_3^2\Omega_n
\nonumber\\
&&\qquad\qquad\qquad   + K_sK_t\Omega_n)\bigr]\quad,
\label{eq:3.4d}\\
D_n({\bf p})&=&\frac{2\pi T}{N}\,\bigl[-a^2K_3 + ab(K_s+K_t) - b(bK_3 + 
                 K_3^2\Omega_n 
\nonumber\\
&&\qquad\qquad\qquad - K_sK_t\Omega_n)\bigr]\quad.
\label{eq:3.4e}
\eea
\eml%

\subsection{Perturbation theory to one-loop order}
\label{subsec:III.B}

We now proceed to perform a one-loop renormalization of the theory.
We do this by renormalizing the two-point vertex ${^{ij}_{rs}M}_{12,34}$,
Eqs.\ (\ref{eqs:3.1}). This procedures proves the renormalizability of the
theory to one-loop order, i.e. it makes sure that no coupling
constants in addition to the ones present in the bare theory
are generated under renormalization. We also need to determine the
wavefunction renormalization. This we do by considering the one-point
vertex function $\Gamma^{(1)} = \langle{\hat Q}\rangle^{-1}$.

\subsubsection{One-point vertex}
\label{subsubsec:III.B.1}

Let us first consider the one-point propagator $\langle{\hat Q}\rangle$.
To one-loop order, the only diagram that contributes is shown in 
Fig.\ \ref{fig:2}.
\begin{figure}[ht]
\vskip 0.2cm
\epsfxsize=4.00cm
\centerline{\epsffile{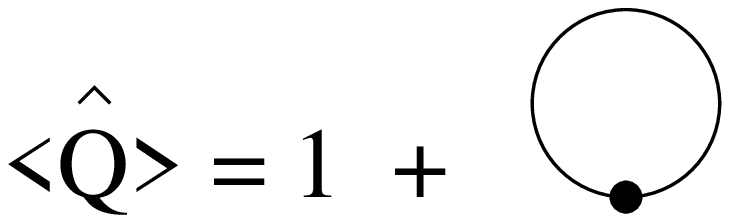}}
\vskip 0.5cm
\caption{Perturbation theory for $\langle{\hat Q}\rangle$ to one-loop order.}
\label{fig:2}
\end{figure} 
In spin-quaternion space, there are two nonvanishing matrix elements of
$\langle{\hat Q}\rangle$, viz. $\langle{^0_0Q}\rangle$ and
$\langle{^3_3Q}\rangle$. These expectation values are diagonal in
both frequency and replica space. Their inverses constitute one-point vertex 
functions that we denote by $\Gamma_0^{(1)}(\Omega_n)$ and
$\Gamma_3^{(1)}(\Omega_n)$, respectively. A simple calculation using the
results of Sec.\ \ref{subsec:III.A} yields
\bml
\label{eqs:3.5}
\bea
\Gamma_0^{(1)}(\Omega_n)&=&1 + \frac{1}{8}\,[I_1^s(\Omega_n) + I_1^t(\Omega_n)]
   \quad,
\label{eq:3.5a}\\
\Gamma_3^{(1)}(\Omega_n)&=&1 + \frac{1}{4}\,I_1^3(\Omega_n)\quad.
\label{eq:3.5b}
\eea
\eml%
Here we have defined the integrals
\bml
\label{eqs:3.6}
\bea
I_1^{s,t}(\Omega_n)&=&\frac{1}{V}\sum_{\bf p}\sum_{l=n}^{\infty} 
   C_l^{s,t}({\bf p})\quad,
\label{eq:3.6a}\\
I_1^{3}(\Omega_n)&=&\frac{1}{V}\sum_{\bf p}\sum_{l=n}^{\infty} 
   D_l({\bf p})\quad,
\label{eq:3.6b}
\eea
\eml%

\subsubsection{Two-point vertex}
\label{subsubsec:III.B.2}

We now turn to the two-point vertex $\Gamma^{(2)}$, whose Gaussian
approximation is given by Eqs.\ (\ref{eqs:3.1}). To one-loop order,
we write
\be
{^{ij}_{rs}\Gamma}^{(2)}_{12,34}({\bf p}) = {^{ij}_{rs}M}_{12,34}({\bf p})
   + {^{ij}_{rs}(\delta M)}_{12,34}({\bf p}) \quad.
\label{eq:3.7}
\ee
There are two topologically distinct diagrammatic contributions to
$\delta M$, which are shown in Fig.\ \ref{fig:3}. 
\begin{figure}[ht]
\vskip 0.3cm
\epsfxsize=4.5cm
\centerline{\epsffile{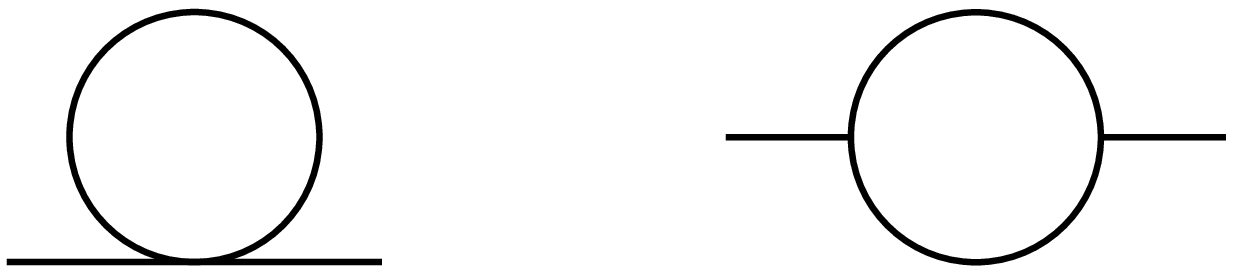}}
\vskip 0.5cm
\caption{One-loop contributions to the two-point vertex.}
\label{fig:3}
\end{figure}
They arise from quartic terms, i.e. terms of $O(q^4)$, and cubic terms,
i.e. terms of $O(q^3)$, respectively, in the expansion of the action in
powers of $q$. An evaluation of the diagrams is straightforward but very
tedious, since the two topological structures can be dressed in many ways
with the various indices carried by the $q$-field. The calculation reveals
that the one-loop contributions can be grouped into three distinct classes:
(1) Quartic contributions that are logarithmically divergent as 
$\Omega_n\rightarrow 0$ in $d=2$, (2) Cubic contributions that have the same
degree of divergence, and (3) contributions, both quartic and cubic, which
individually diverge more strongly (`superdivergent terms'), 
but combine to yield again terms that are only logarithmically divergent.
Denoting the contributions of these three classes to the one-loop
renormalization of $\Gamma^{(2)}$ by $\delta M^{(4)}$, $\delta M^{(3)}$,
and $\delta M^{(sd)}$, respectively, we find for the first of these classes
\bml
\label{eqs:3.8}
\bea
{^{ij}_{rs}(}\delta M)^{(4)}_{12,34}({\bf p})&=&\delta_{13}\,\delta_{24}\,
   \left\{\delta_{rs}\delta_{ij}\frac{1}{8G}\,({\bf p}^2 + GH\Omega_N)\right.
\nonumber\\
&&\qquad\qquad\qquad\times [I_1^{(s)}(\Omega_N) + I_1^{(t)}(\Omega_N)]
\nonumber\\
&&\hskip -15pt + \delta_{rs}\delta_{ij}\,\frac{1}{8G_3}\,
   ({\bf p}^2 + G_3H_3\Omega_N)\,2\,I_1^{(3)}(\Omega_N)
\nonumber\\
&&\hskip -15pt + m_{rs,ij}\,\frac{1}{8G}\,({\bf p}^2 + GH\Omega_N)\,2
     I_1^{(3)}(\Omega_N)
\nonumber\\
&&\hskip -65pt \left. +m_{rs,ij}\,\frac{1}{8G_3}\,({\bf p}^2 + G_3H_3\Omega_N)\,
  [I_1^{(s)}(\Omega_N) + I_1^{(t)}(\Omega_N)]
   \right\} 
\nonumber\\
&&\hskip -75pt -\delta_{1-2,3-4}\delta_{\alpha_1\alpha_2}
      \delta_{\alpha_1\alpha_3}\frac{\pi}{4}\,T\,\biggl\{\delta_{rs}\delta_{ij}
       \bigl[(K_s + K_t)\,J(\Omega_N)
\nonumber\\
&&\qquad\qquad\qquad+2K_3J^{(3)}(\Omega_N)\bigr]
\nonumber\\
&&\hskip -65pt + m_{rs,ij}\bigl[2K_3 J(\Omega_N)
   + (K_s +K_t)\,J^{(3)}(\Omega_N)\bigr]\biggr\}\ ,
\label{eq:3.8a}
\eea
with $I_1^{(s,t,3)}$ given by Eqs.\ (\ref{eqs:3.6}),
\bea
J(\Omega_N) = \frac{1}{V}\sum_{\bf p} A_N({\bf p})\quad,
\label{eq:3.8b}\\
J^{(3)}(\Omega_N) = \frac{1}{V}\sum_{\bf p} B_N({\bf p})\quad,
\label{eq:3.8c}
\eea
and $N$ an external frequency (e.g., $N=n_1-n_2$).
\eml%
For the second class we obtain
\bml
\label{eqs:3.9}
\bea
{^{ij}_{rs}(}\delta M)^{(3)}_{12,34}({\bf p})&=&-\delta_{1-2,3-4}
   \delta_{\alpha_1\alpha_2}\delta_{\alpha_1\alpha_3}(\pi T)^2\biggl\{
   \delta_{rs}\delta_{ij}
\nonumber\\
&&\hskip -30pt\times\bigr[ (\delta_{i0}K_s^2 + \delta_{i3}K_sK_t + K_3^2)\,
                     I_4^{(s)}(\Omega_N)
\nonumber\\ 
&&\hskip -22pt + (\delta_{i3}K_t^2 + \delta_{i0}K_sK_t + K_3^2)
      \,I_4^{(t)}(\Omega_N)
\nonumber\\
&&\hskip -22pt + (\delta_{i0}K_sK_3 + K_sK_3 + \delta_{i3}K_tK_3)
      \,I_5^{(s)}(\Omega_N)
\nonumber\\
&&\hskip -22pt + (\delta_{i0}K_sK_3 + \delta_{i3}K_tK_3 + K_tK_3)
      \,I_5^{(t)}(\Omega_N)\bigr]
\nonumber\\
&&\hskip -60pt + m_{rs,ij}\bigl[(\delta_{i0}K_s^2 + \delta_{j3}K_sK_t + K_3^2)
   \,I_5^{(s)}(\Omega_N)
\nonumber\\
&&\hskip -22pt + (\delta_{i3}K_t^2 + \delta_{i0}K_sK_t + K_3^2)
          \,I_5^{(t)}(\Omega_N)
\nonumber\\
&&\hskip -22pt + (\delta_{i0}K_sK_3 + K_sK_3 + \delta_{i3}K_tK_3)
          \,I_4^{(s)}(\Omega_N)
\nonumber\\
&&\hskip -22pt + (\delta_{i0}K_sK_3 + \delta_{i3}K_tK_3 + K_tK_3)
          \,I_4^{(t)}(\Omega_N)\bigr]\biggr\}
\nonumber\\
\label{eq:3.9a}
\eea
Here we have defined integrals
\bea
I_4^{(s,t)}(\Omega_N)&=&\frac{1}{V}\sum_{\bf p}\sum_{l=N}^{\infty}\,
    \bigl[A_l({\bf p})^2 + B_l({\bf p})^2
\nonumber\\
&&\hskip -20pt + A_l({\bf p})\,l\,C_l^{(s,t)}({\bf p})
   + B_l({\bf p})\,l\,D_l({\bf p})\bigr]\ ,
\label{eq:3.9b}\\
I_5^{(s,t)}(\Omega_N)&=&\frac{1}{V}\sum_{\bf p}\sum_{l=N}^{\infty}\,
   \bigl[2\,A_l({\bf p})\,B_l({\bf p})
\nonumber\\
&&\hskip -20pt + B_l({\bf p})\,l\,C_l^{(s,t)}({\bf p})
   + A_l({\bf p})\,l\,D_l({\bf p})\bigr]\ ,
\label{eq:3.9c}
\eea
\eml%
Finally, for the third class we have
\bml
\label{eqs:3.10}
\bea
{^{ij}_{rs}(}\delta M)^{(sd)}_{12,34}({\bf p})&=&\delta_{13}\delta_{24}\,
   \frac{1}{8}\,\biggl\{\delta_{rs}\delta_{ij}\bigl[I_2^{(s)}({\bf p},\Omega_N)
\nonumber\\
&&\qquad\qquad\qquad\qquad + I_2^{(t)}({\bf p},\Omega_N)\bigr]
\nonumber\\
&&\hskip -35pt + m_{rs,ij}\,\bigl[I_3^{(s)}({\bf p},\Omega_N)
                           + I_3^{(t)}({\bf p},\Omega_N)\bigr]\biggr\}\ ,
\label{eq:3.10a}
\eea
with integrals
\bea
I_2^{(s,t)}({\bf p},\Omega_N)&=&\frac{1}{V}\sum_{\bf k}\sum_{l=1}^{\infty}
   \biggl\{a_l({\bf k})\,C_l^{(s,t)}({\bf k}) + b_l({\bf k})\,D_l({\bf k})
\nonumber\\
&&\hskip -50pt + 2\pi\,T\,K_{s,t}\,A_l({\bf k}) + 2\pi\,T\,K_3\,B_l({\bf k})
\nonumber\\
&&\hskip -50pt -(2\pi\,T)^2K_{s,t}^2\,l\,A_{l+N}({\bf k}+{\bf p})\bigl[
     A_l({\bf k}) + l\,C_l^{(s,t)}({\bf k})\bigr]
\nonumber\\
&&\hskip -50pt -(2\pi\,T)^2K_sK_t\,l\,B_{l+N}({\bf k}+{\bf p})\bigl[
     B_l({\bf k}) + l\,D_l({\bf k})\bigr]
\nonumber\\
&&\hskip -50pt -(2\pi\,T)^2K_3^2\,l\,A_{l+N}({\bf k}+{\bf p})\bigl[
     A_l({\bf k}) + l\,C_l^{(s,t)}({\bf k})\bigr]
\nonumber\\
&&\hskip -50pt -(2\pi\,T)^2K_3^2\,l\,B_{l+N}({\bf k}+{\bf p})\bigl[
     B_l({\bf k}) + l\,D_l({\bf k})\bigr]
\nonumber\\
&&\hskip -50pt -(2\pi\,T)^2 2K_3K_{s,t}\,l\,A_{l+N}({\bf k}+{\bf p})\bigl[
     2B_l({\bf k}) + l\,D_l({\bf k})\bigr]
\nonumber\\
&&\hskip -50pt -(2\pi\,T)^2 2K_3K_{s,t}\,l\,B_{l+N}({\bf k}+{\bf p})\,
     l\,C_l^{(s,t)}({\bf k})\biggr\}\quad,
\nonumber\\
\label{eq:3.10b}\\
I_3^{(s,t)}({\bf p},\Omega_N)&=&\frac{1}{V}\sum_{\bf k}\sum_{l=1}^{\infty}
   \biggl\{b_l({\bf k})\,C_l^{(s,t)}({\bf k}) + a_l({\bf k})\,D_l({\bf k})
\nonumber\\
&&\hskip -50pt + 2\pi\,T\,K_{s,t}\,B_l({\bf k}) + 2\pi\,T\,K_3\,A_l({\bf k})
\nonumber\\
&&\hskip -50pt -(2\pi\,T)^2K_{s,t}^2\,l\,B_{l+N}({\bf k}+{\bf p})\bigl[
     A_l({\bf k}) + l\,C_l^{(s,t)}({\bf k})\bigr]
\nonumber\\
&&\hskip -50pt -(2\pi\,T)^2K_sK_t\,l\,A_{l+N}({\bf k}+{\bf p})\bigl[
     B_l({\bf k}) + l\,D_l({\bf k})\bigr]
\nonumber\\
&&\hskip -50pt -(2\pi\,T)^2K_3^2\,l\,B_{l+N}({\bf k}+{\bf p})\bigl[
     A_l({\bf k}) + l\,C_l^{(s,t)}({\bf k})\bigr]
\nonumber\\
&&\hskip -50pt -(2\pi\,T)^2K_3^2\,l\,A_{l+N}({\bf k}+{\bf p})\bigl[
     B_l({\bf k}) + l\,D_l({\bf k})\bigr]
\nonumber\\
&&\hskip -50pt -(2\pi\,T)^2 2K_3K_{s,t}\,l\,A_{l+N}({\bf k}+{\bf p})\bigl[
     A_l({\bf k}) + l\,C_l^{(s,t)}({\bf k})\bigr]
\nonumber\\
&&\hskip -50pt -(2\pi\,T)^2 2K_3K_{s,t}\,l\,B_{l+N}({\bf k}+{\bf p})\bigl[
     B_l({\bf k}) + l\,D_l({\bf k})\bigr]\biggr\}\ .
\nonumber\\
\label{eq:3.10c}
\eea
\eml%
As a check, we consider the superdivergent contributions at zero external
frequency and wavenumber. Although all of the individual terms are linearly
divergent, an explicit calculation yields
\bml
\label{eqs:3.11}
\be
I_2^{(s)}(0,0) + I_2^{(t)}(0,0) = 0\quad, 
\label{eq:3.11a}
\ee
and the same for $I_3$. Therefore,
\be
{^{ij}_{rs}(}\delta M)^{(sd)}_{12,34}({\bf p}) = O({\bf p}^2,\Omega_N)\quad,
\label{eq:3.11b}
\ee
\eml%
with coefficients of ${\bf p}^2$ and $\Omega_N$ that are only logarithmically
divergent. The cancellation of the superdivergencies thus holds as expected
(and required by, e.g., particle number conservation and the renormalizability
of the theory.)

\subsection{Expansion to linear order in the magnetic coupling constants}
\label{subsec:III.C}

As is clear from the preceding subsection, the complete one-loop renormalization
of our model is rather complicated. While it is certainly possible to determine
the renormalization group (RG) flow equations from our perturbative results,
it would not be easy to analyze them for fixed points. At this point we
therefore take a less general approach that is based on the following physical
considerations. We are interested in a phase transition from a metallic
magnetic phase to an insulating magnetic phase. Physically, we expect the
magnetization to be noncritical at such
a transition. The simplest possible scenario is then a fixed point where
the renormalized values of the `magnetic' coupling constants, $1/G_3$,
$H_3$, and $K_3$, are all zero. (More complicated possibilities we will
come back to in Sec.\ \ref{sec:IV} below.) This means that the unversality
class of this transition is the same as that for the transition from a
paramagnetic metal to a paramagnetic insulator in the presence of an
external magnetic field.\cite{MF_Footnote} We 
can check this scenario by expanding
to linear order in the three magnetic coupling constants and investigate
the perturbative stability of the nonmagnetic fixed point.

Accordingly, we expand the results of the previous subsection, expressing
the result in the form of corrections to the magnetic coupling constants.
We use dimensional regularization, i.e. we
perform the integrals in $d=2+\epsilon$ to leading order in $1/\epsilon$. 
We find
for the correction to $1/G_3$ to linear order in $1/G_3$, $H_3$, and $K_3$,
\bml
\label{eqs:3.12}
\bea
\delta(1/G_3)&=&\frac{2}{\epsilon}\left[\frac{G}{16 G_3}\,f_{11}(K_s/H,K_t/H)
       \right.
\nonumber\\
&&\hskip -30pt   + \frac{H_3}{8H}\,f_{12}(K_s/H,K_t/H)
   + \frac{K_3}{4H}\,f_{13}(K_s/H,K_t/H)\biggr]\ .
\nonumber\\
\label{eq:3.12a}
\eea
Here we have defined the functions
\bea
f_{11}(x,y)&=&g_{11}(x) + g_{11}(y) - 2\,\frac{L_x - L_y}{x-y}
\nonumber\\
&&\hskip 20pt  + 2\,\frac{xy}{x-y}\,\left[h_{11}(x) - h_{11}(y)\right]\quad,
\label{eq:3.12b}\\
f_{12}(x,y)&=&g_{12}(x) + g_{12}(y) + \frac{L_x - L_y}{x-y}
\nonumber\\
&&\hskip 20pt  + \frac{xy}{x-y}\,\left[h_{12}(x) - h_{12}(y)\right]\quad,
\label{eq:3.12c}\\
f_{13}(x,y)&=&\frac{-1}{x-y}\,\left(\frac{1}{x}\,L_x - \frac{1}{y}\,L_y\right)
   \quad,
\label{eq:3.12d}
\eea
in terms of
\bea
g_{11}(x)&=&\frac{-6}{x} - 2 + \frac{6}{x}\,\left(\frac{1}{x} + 1\right)\,
   \ln (1+x)\quad,
\label{eq:3.12e}\\
h_{11}(x)&=&\frac{-2}{x^2} - \frac{2}{x} + \frac{1}{x^2}\,\left(\frac{2}{x}
   + 3\right)\,\ln (1+x)\quad,
\label{eq:3.12f}\\
g_{12}(x)&=&\frac{1}{x}\,\left[3 - \left(\frac{3}{x} + 2\right)\,\ln (1+x)
   \right]\quad,
\label{eq:3.12g}\\
h_{12}(x)&=&\frac{1}{x^2}\,\left[2 - \left(\frac{2}{x} + 1\right)\,\ln (1+x)
   \right]\quad,
\label{eq:3.12h}
\eea
\eml%
and $L_x = \ln (1+x)$, $L_y = \ln (1+y)$. 

Similarly, the correction to
$H_3$ is
\bml
\label{eqs:3.13}
\bea
\delta H_3&=&\frac{G}{4\epsilon}\,\left[\frac{G H}{G_3}\,f_{21}(K_s/H,K_t/H)
   \right.
\nonumber\\
&&\hskip -20pt  + H_3\,f_{22}(K_s/H,K_t/H)
   + K_3\,f_{23}(K_s/H,K_t/H)\biggr]\ .
\nonumber\\
\label{eq:3.13a}
\eea
with
\bea
f_{21}(x,y)&=&g_{21}(x) + g_{21}(y) - \frac{1+x+y}{x-y}\,\left(L_x-L_y\right)
\nonumber\\
&&\hskip 20pt  - \frac{xy}{x-y}\,\left[h_{21}(x) - h_{21}(y)\right]\quad,
\label{eq:3.13b}\\
f_{22}(x,y)&=&g_{22}(x) + g_{22}(y) + \frac{L_x - L_y}{x-y}
\nonumber\\
&&\hskip 20pt  + \frac{xy}{x-y}\,\left[h_{21}(x) - h_{21}(y)\right]\quad,
\label{eq:3.13c}\\
f_{23}(x,y)&=&g_{23}(x) + g_{23}(y) + \frac{L_x - L_y}{x-y}
\nonumber\\
&&\hskip 20pt  + \frac{xy}{x-y}\,\left[h_{21}(x) - h_{21}(y)\right]\quad,
\label{eq:3.13d}
\eea
in terms of
\bea
g_{21}(x)&=&1 + \frac{x}{2} - \left(\frac{1}{x} + 1\right)\,\ln (1+x)\quad,
\label{eq:3.13e}\\
h_{21}(x)&=&\frac{1}{x^2}\,\ln (1+x)\quad,
\label{eq:3.13f}\\
g_{22}(x)&=&-1 + \left(\frac{1}{x} + 1\right)\,\ln (1+x)\quad,
\label{eq:3.13g}\\
g_{23}(x)&=&\frac{-1}{2} + \frac{1}{x}\,\ln (1+x)\quad.
\label{eq:3.13h}
\eea
\eml%

Finally, for the correction to $K_3$ we obtain
\bml
\label{eqs:3.14}
\bea
\delta K_3&=&\frac{G}{8\epsilon}\,\left[\frac{G H}{G_3}\,f_{31}(K_s/H,K_t/H)
        \right.
\nonumber\\
&&\hskip -20pt  + H_3\,f_{32}(K_s/H,K_t/H)
   + K_3\,f_{33}(K_s/H,K_t/H)\biggr]\ .
\nonumber\\
\label{eq:3.14a}
\eea
with
\bea
f_{31}(x,y)&=&x + y + \frac{x+y}{x-y}\,\left[h_{31}(x) - h_{31}(y)\right]\quad,
\label{eq:3.14b}\\
f_{32}(x,y)&=&g_{32}(x) + g_{32}(y) + \frac{y}{x}\,L_x + \frac{x}{y}\,L_y
\nonumber\\
&&\hskip 20pt  + \frac{(x+y)^2}{x-y}\,\left[h_{32}(x) - h_{32}(y)\right]\quad,
\label{eq:3.14c}\\
f_{33}(x,y)&=&g_{33}(x) + g_{33}(y) + \frac{y}{x}\,L_x + \frac{x}{y}\,L_y
\nonumber\\
&&\hskip 20pt  + \frac{(x+y)^2}{x-y}\,\left[h_{32}(x) - h_{32}(y)\right]\quad,
\label{eq:3.14d}
\eea
in terms of
\bea
h_{31}(x)&=&-2(1+x)\,\ln (1+x)\quad,
\label{eq:3.14e}\\
g_{32}(x)&=&-2x + \ln (1+x)\quad,
\label{eq:3.14f}\\
h_{32}(x)&=&\frac{1}{x}\,\ln (1+x)\quad,
\label{eq:3.14g}\\
g_{33}(x)&=&1 + 3\ln (1+x)\quad.
\label{eq:3.14h}
\eea
\eml%

An inspection of the integrals in Sec.\ \ref{subsec:III.B} further shows
that all corrections to the remaining coupling constants, $G$, $H$, and 
$K_{s,t}$, to the extent that they depend on the magnetic coupling constants,
are at least quadratic in the latter and hence can be neglected for our
purposes. The `nonmagnetic' one-loop corrections are well known,\cite{R}
and we do not write them down again.

We also note that $1/G_3 \neq 0$ is sufficient to generate nonzero values
of $H_3$ and $K_3$ in perturbation theory, even if these coupling constants
were not present in the bare action. This is the reason why we have included
them in Eq.\ (\ref{eq:2.18a}).

\subsection{Renormalization group flow equations}
\label{subsec:III.D}

We now perform a RG analysis of our perturbation theory. We define renormalized
coupling constants $g_3$, $h_3$, and $k_3$ by
\bml
\label{eqs:3.15}
\bea
G_3&=&\kappa^{-\epsilon}Z_{g_3}g_3\quad,
\label{eq:3.15a}\\
H_3&=&Z_{h_3}h_3\quad,
\label{eq:3.15b}\\
K_3&=&Z_{k_3}k_3\quad,
\label{eq:3.15c}
\eea
\eml%
where the $Z$ are renormalization constants, and $\kappa$ is the arbitrary RG
momentum scale.\cite{ZJ} We further define a two-point vertex function
$\Gamma^{(2)}_3$ as the `magnetic piece' of the general vertex $\Gamma^{(2)}$
defined in Eq.\ (\ref{eq:3.7}), i.e. the parts that are proportional to
$1/G_3$, $H_3$, and $K_3$. From 
Eqs.\ (\ref{eqs:3.1},\ref{eq:3.7},\ref{eqs:3.8}-\ref{eqs:3.10}) we have
\bea
\Gamma^{(2)}_3({\bf p},\Omega)&=&\left(\frac{1}{G_3} + \delta(1/G_3)\right)\,
   {\bf p}^2 + \left(H_3 + \delta H_3\right)\,\Omega
\nonumber\\
&&\hskip 20pt   + \left(K_3 + \delta K_3\right)\,\Omega\quad.
\label{eq:3.16}
\eea
The renormalization constants can then be determined from the renormalization
statement
\be
\Gamma^{(2)}_{3,\rm R}({\bf p},\Omega;g_3,h_3,k_3;\kappa) 
   = Z\,\Gamma^{(2)}_3({\bf p},\Omega;G_3,H_3,K_3)\quad,
\label{eq:3.17}
\ee
where $\Gamma^{(2)}_{3,\rm R}$ is the renormalized counterpart of
$\Gamma^{(2)}_{3}$, and $Z$ is the wavefunction renormalization. In
our notation, we suppress the dependence of the vertex functions on
the remaining coupling constants, $G$, $H$, $K_{s,t}$, and their
renormalized counterparts.

It is {\it a priori} not clear that a single wavefunction renormalization
constant will suffice. Indeed, the existence of two distinct one-point
vertex functions, Eqs.\ (\ref{eqs:3.5}), one being related to the density
of states and the other to the magnetization, might suggest that one needs
at least two. However, as mentioned in Sec.\ \ref{subsec:III.C} above,
we do not expect the magnetization to display leading critical behavior
at the phase transition we are interested in, despite the fact that the
magnetization has nonanalytic contributions in perturbation
theory. We therefore expect the only wavefunction renormalization to
be the one related the vertex $\Gamma_0^{(1)}$,
\be
\Gamma^{(1)}_{1,\rm R}(\Omega;g_3,h_3,k_3;\kappa) 
    = Z\,\Gamma^{(1)}_1(\Omega;G_3,H_3,K_3)\quad.
\label{eq:3.18}
\ee
To linear order in $1/g_3$, $h_3$, and $k_3$, $Z$ is given by the
wavefunction renormalization for nonmagnetic electrons in an external
magnetic field,\cite{R}
\be
Z = 1 - \frac{g}{4\epsilon}\,\left(l_s + l_t\right)\quad,
\label{eq:3.19}
\ee
where $l_{s,t} = \ln (1 + \gamma_{s,t})$, with $\gamma_{s,t}\equiv k_{s,t}/h$ 
the renormalized counterparts of $K_{s,t}/H$. Our perturbative calculation
of $\Gamma^{(2)}_3$ is then sufficient to determine the remaining
renormalization constants. Using minimal subtraction, we find
\bml
\label{eqs:3.20}
\bea
Z_{g_3}&=&1 + \frac{g}{8\epsilon}\,\biggl[ f_{11}(\gamma_s,\gamma_t)
            - 2(l_s + l_t) + 2\,\frac{g_3h_3}{gh}\,f_{12}(\gamma_s,\gamma_t)
\nonumber\\
&&\hskip 50pt + 4\,\frac{g_3k_3}{gh}\,f_{13}(\gamma_s,\gamma_t)\biggr]\quad,
\label{eq:3.20a}\\
Z_{h_3}&=&1 + \frac{g}{4\epsilon}\,\biggl[ l_s + l_t - f_{22}(\gamma_s,\gamma_t)
            - \frac{gh}{g_3h_3}\,f_{21}(\gamma_s,\gamma_t)
\nonumber\\
&&\hskip 50pt - \frac{k_3}{h_3}\,f_{23}(\gamma_s,\gamma_t)\biggr]\quad,
\label{eq:3.20b}\\
Z_{k_3}&=&1 + \frac{g}{8\epsilon}\,\biggl[2(l_s + l_t) 
   - f_{33}(\gamma_s,\gamma_t) - \frac{h_3}{k_3}\,f_{32}(\gamma_s,\gamma_t)
\nonumber\\
&&\hskip 50pt - \frac{gh}{g_3k_3}\,f_{13}(\gamma_s,\gamma_t)\biggr]\quad.
\label{eq:3.20c}
\eea
\eml%

From Eqs.\ (\ref{eqs:3.20}) and (\ref{eqs:3.15}) it is now easy to 
determine the RG flow equations for the magnetic coupling constants.
Our parameter space is spanned by $\mu = (g,h,\gamma_s,\gamma_t,g_3,h_3,k_3)$,
and our approximations are valid only in the vicinity of the
fixed point (FP) $\mu^* = (g^*,h^*,\gamma_s^*,\gamma_t^*,g_3^*,h_3^*,k_3^*)$,
with $1/g_3^*=h_3^*=k_3^*=0$, and $g^*$, $h^*$, $\gamma_s^*$, and $\gamma_t^*$
the FP values of these coupling constants for the magnetic-field universality
class of nonmagnetic electrons.\cite{R} We therefore immediately linearize
about this FP. With $\beta_3 \equiv 1/g_3$, and $\ell \equiv 1/\kappa$ the
RG length scale, we find
\bml
\label{eqs:3.21}
\bea
\frac{d\beta_3}{d\ell}&=&\left(\epsilon - \frac{g^*}{8}\,\left[f_{11}^*
   - 2(l_s^* + l_t^*)\right]\right)\,\beta_3 - \frac{f_{12}^*}{4h^*}\,h_3
\nonumber\\
&&\hskip 50pt   - \frac{f_{13}^*}{2h^*}\,k_3\quad,
\label{eq:3.21a}\\
\frac{dh_3}{d\ell}&=&\frac{-1}{4}\,(g^*)^2\,h^*\,f_{21}^*\,\beta_3
   + \frac{g^*}{4}\,\left[l_s^* + l_t^* - f_{22}^*\right]
\nonumber\\
&&\hskip 50pt   - \frac{g^*}{4}\,f_{23}^*\,k_3\quad,
\label{eq:3.21b}\\
\frac{dk_3}{d\ell}&=&\frac{-1}{8}\,(g^*)^2\,h^*\,f_{31}^*\,\beta_3
   - \frac{g^*}{8}\,f_{32}^*\,h_3 
\nonumber\\
&&\hskip 40pt + \frac{g^*}{8}\,\left[2(l_s^* + l_t^*) - f_{33}^*\right]\,k_3
   \quad,
\label{eq:3.21c}
\eea
\eml%
where $f_{11}^* \equiv f_{11}(\gamma_s^*,\gamma_t^*)$, etc.

The fixed point values that enter Eqs.\ (\ref{eqs:3.21}) depend on whether
we consider the long-ranged Coulomb interaction between the electrons, or
a short-ranged model interaction. We consider here the former, more
realistic, case. Then we have\cite{R}
\be
g^* = 2\epsilon/(1-\ln 2)\quad,\quad \gamma_t^* = -\gamma_s^* = 1\quad,
   \quad l_s^* = -2/\epsilon\quad.
\label{eq:3.22}
\ee
With this input, we obtain the following eigenvalues for the linearized
flow equations, Eqs.\ (\ref{eqs:3.21}),
\bml
\label{eqs:3.23}
\bea
\lambda_1&=&-\epsilon/2(1-\ln 2) + O(\epsilon^2) < 0\quad,
\label{eq:3.23a}\\
\lambda_2&=&-1/(1-\ln 2) + O(\epsilon) < 0 \quad,
\label{eq:3.23b}\\
\lambda_3&=&-\,\frac{3\ln 2 - 2}{1 - \ln 2}\,\epsilon + O(\epsilon^2) < 0\quad.
\label{eq:3.23c}
\eea
\eml%
We see that all three eigenvalues are negative, and the fixed point is
therefore stable.

\subsection{Critical behavior}
\label{subsec:III.E}

As we have seen in the previous subsection, the critical fixed point for
the transition under consideration is the same as the one found before
for the metal-insulator transition of nonmagnetic electrons in the
presence of an external magnetic field.\cite{MF_Footnote} 
The asymptotic critical behavior
is therefore also the same. Choosing the correlation length exponent
$\nu$, the critical exponent for the density of states $\beta$, and
the dynamical critical exponent $z$ as the three independent exponents,
we thus have\cite{R}
\bml
\label{eqs:3.24}
\bea
\nu&=&1/\epsilon + O(1)\quad,
\label{eq:3.24a}\\
\beta&=&1/2\epsilon (1 - \ln 2)\quad,
\label{eq:3.24b}\\
z&=&d \quad.
\label{eq:3.24c}
\eea
The critical exponent for the conductivity, $s=\nu (d-2)$, is
\be
s = 1 + O(\epsilon)\quad.
\label{eq:3.24d}
\ee
\eml%

In contrast to the asymptotic critical behavior, the corrections to scaling
are different from any previously studied universality class for 
metal-insulator transitions. The reason for this is the presence of the
three irrelevant operators $1/g_3$, $h_3$, and $k_3$ in our model. 
We will not go through a complete analysis of the corrections to scaling
here, but only mention that they lead to a nonanalyticity in the magnetization
as one crosses the metal-insulator transition, even though the
magnetization is not critical. To see this, we recall that the magnetization
is proportional to a frequency integral over the inverse of the
one-point vertex $\Gamma_3^{(1)}$, see Eq.\ (\ref{eq:3.5b}) above and
Eq.\ (2.7c) in (I). The extra frequency integration makes the integral
finite for all $d>0$, and the one-loop contribution to the magnetization
is simply proportional to $1/g_3$, $h_3$, and $k_3$. Since $\lambda_3$
has the smallest absolute value of the three negative eigenvalues given
in Eq.\ (\ref{eqs:3.23}), the magnetization at $T=0$ behaves like
\bml
\label{eqs:3.25}
\be
M(t,T=0) \propto {\rm const.} + t^{-\nu\lambda_3}\quad,
\label{eq:3.25a}
\ee
where $t$ is the dimensionless distance from the critical point. At
criticality as a function of temperature we have
\be
M(t=0,T) \propto {\rm const.} + T^{-\lambda_3/z}\quad.
\label{eq:3.25b}
\ee
\eml%
Putting $\epsilon = 1$ in our one-loop approximation yields
$-\lambda_3/z = 0.086\ldots$. Our theory thus predicts that the
metal-insulator transition is reflected in the magnetization in the form
of a very slow temperature dependence.

More generally, the existence of very slow corrections to scaling indicates
that it will be very difficult, if not impossible, to observe the true
asymptotic critical behavior at the ferromagnetic metal-insulator transition.
If we extrapolate our one-loop results to $d=3$ by putting $\epsilon=1$,
we have $\nu\lambda_3 \approx -0.26$. This mean that in order to obtain the
critical exponents with an accuracy of $10\%$ one needs to be within about $0.01\%$
of the critical point, $t\alt 10^{-4}$. This is not achievable for any
metal-insulator transition observed so far.\cite{R} Any observed critical 
behavior at larger values of $t$ will yield effective exponents that contain
contributions from the dominant irrelevant scaling variables. We note in
passing that the same conclusion holds for the Anderson-Mott transition of
paramagnetic electrons in an external magnetic field.\cite{MF_Footnote}

\section{Discussion}
\label{sec:IV}

Our chief result is the prediction that the metal-insulator transition from
a ferromagnetic metal to a ferromagnetic insulator is in the same universality
class as the one from a paramagnetic metal to a paramagnetic insulator in the
presence of an external magnetic field. It is important to note that this
statement holds independent of what the actual critical exponents, which we
can determine only to lowest order in a $2+\epsilon$ expansion, are in 
three-dimensional systems. It is also independent of the fact that
we have considered
explicitly only the perturbative stability of the nonmagnetic fixed point. In
principle, the full flow equations that follow from the one-loop calculation
in Sec.\ \ref{subsec:III.B} could contain other critical fixed points. This
is a question that remains to be investigated; here we just mention a
possible scenario. 

In both the magnetic field and ferromagnetic material cases a different
universality class for the MIT is easy to envisage. First note that the
existence of two conductivities, $\sigma^{\pm}$ (cf. Eqs.\ (\ref{eqs:2.16})), 
or equivalently,
two diffusivities, just reflects the fact that either a magnetic field or a
finite spontaneous magnetization leads to a splitting of the energy band. The
subband with fewer (more) electrons, that have spins aligned in (opposite to)
the direction of 
the magnetic field or spontaneous magnetization is called the
minority (majority) subband. If the magnetic energy scale is large compared to
other interaction energy scales and comparable to the Fermi energy, then the
two subbands are well separated. A polarization scenario for the MIT is
that the minority subband carriers become localized first and then
act as a static random field for the majority mobile carriers.\cite{FR} 
In this scenario, the MIT
occurs when the carriers in the majority band become localized. In this case
the MIT is one for spin-polarized, or effectively spinless, electrons. This is
mathematically described by the so-called singlet-only or magnetic impurity
universality class that was discussed in Ref.\ \onlinecite{Cetal}. 
One might thus expect a multicritical point
separating the magnetic field universality class, which was discussed above
and is relevant for small values of the magnetization or the magnetic field,
from the singlet-only or
polarization universality class at large values of the magnetization. 
Experimentally, such a multicritical point could be probed by
increasing the magnetic field in the case of an MIT in an external magnetic
field, or by effectively increasing the triplet interaction for a
spontaneously magnetized system. Theoretically, it remains to be seen
whether such a behavior is
described by our complete flow equations. This point will be investigated in
a future publication.

In any event, it would be very
interesting to compare experiments on a ferromagnetic metal-to-insulator
transition, which has not been studied so far, with the existing results for
nonmagnetic systems in a magnetic field.\cite{R} The equivalence of the two
universality classes also leads to the conclusion that the existing theory for
the nonmagnetic transition in a magnetic field is incomplete since it misses
important corrections to scaling.\cite{MF_Footnote}

From a theoretical point of view, this result is {\it a priori} rather 
surprising.
The critical behavior at the metal-insulator transition, and hence the 
universality
class, is determined by the structure of the soft modes in the system, at least
near two-dimensions. Since ferromagnetism leads to additional soft modes, namely
the Goldstone modes or spin waves, compared to paramagnetic metals, one would
expect the critical behavior to change. The reason why it does not lies in the
fact that the Goldstone modes do not lead to a singular correction to the
conductivity in $d=2$, in contrast to the diffusive soft modes that are also
present in the absence of ferromagnetic long-range order. Since these singular
corrections drive the transition in low dimensions, and since the Goldstone
modes are the only substantial difference between the soft-mode spectra of
ferromagnetic systems and paramagnetic systems in a magnetic field, 
respectively, the fact that
the universality class remains unchanged is at least plausible. 

We finally mention again that one of our motivations for the present study 
had been the observed apparent metal-insulator transition in certain $2-d$ 
electron systems,\cite{2dMIT} which contradicts the results of orthodox 
theories and is not understood. Since it is known that ferromagnetic 
fluctuations enhance the conductivity in $d=2$,\cite{R} it was a plausible 
hypothesis that ferromagnetic long-range order might have an even stronger 
effect and lead to a metallic phase in $d=2$. Our results rule out this 
possibility, at least on a perturbative level.

\acknowledgments
We gratefully acknowledge stimulating discussions with Nick Giordano.
Parts of this work were performed at the Institute for Theoretical Physics
at UC Santa Barbara, and at the Aspen Center for Physics. This work was 
supported by the NSF under grant Nos. DMR-98-70597, DMR-99-75259, 
and PHY94-07194.

\end{document}